\documentclass[letterpaper,11pt]{article}
\begin{document}
\addtolength{\textheight}{+0.18in}
 
\newtheorem{lemma}{Lemma}[section]
\newtheorem{theorem}{Theorem}
\newtheorem{cor}{Corollary}[section]
\newtheorem{prop}{Proposition}[section]
\newcommand{\lf}{\left\lfloor}
\newcommand{\rf}{\right\rfloor}
\newcommand{\dis}{\displaystyle}
\newcommand{\sums}[2]{\sum_{\stackrel{{\scriptstyle {#1}}}{{#2}}}}
\newcommand{\prods}[2]{\prod_{\stackrel{{\scriptstyle {#1}}}{{#2}}}}
\newtheorem{Definition}{Definition}
\newtheorem{corollary}{{Corollary}}
\newcommand{\beq}{\begin{eqnarray*}}
\newcommand{\eeq}{\end{eqnarray*}}
\newcommand{\letab}{\le &}

\newenvironment{tabAlgorithm}[2]{
\setcounter{algorithmLine}{1} \samepage
\begin{tabbing}
999\=\kill #1 \ \ --- \ \ \parbox{3.8in}{\it #2} }{
\end{tabbing}
}
\newcounter{algorithmLine}
\newcommand{\algline}{\\\thealgorithmLine\hfil\>\stepcounter{algorithmLine}}
\newcommand{\algnono}{\\ \>}

\newcommand{\TRUE}{{\bf TRUE}}
\newcommand{\FALSE}{{\bf FALSE}}
\newcommand{\NIL}{{\bf NIL}}
\newcommand{\CURRENT}{{\bf CURRENT}}
\newcommand{\IF}{{\bf IF }\=}
\newcommand{\THEN}{{\bf THEN }\=}
\newcommand{\ELSE}{{\bf ELSE }}
\newcommand{\WHILE}{{\bf WHILE }\=}
\newcommand{\FOR}{{\bf FOR }\=}
\newcommand{\DO}{{\bf DO }\=}
\newcommand{\RETURN}{{\bf RETURN }}
\newcommand{\BREAK}{{\bf BREAK }}

\input epsf
 
\def\prbox{\hfill\rule{1.2ex}{1.2ex}\vspace{.2in}}
 
\newenvironment{proof}{\noindent{\bf Proof }}{\prbox}
 
\def\Reals{\hbox{\rm I\kern-.18em R}}
\def\reals{\Reals}
\def\Complexes{\hbox{\rm C\kern-.43em
       \vrule depth 0ex height 1.4ex width .05em\kern.41em}}
\def\complexes{\Complexes}
 
\def\field{\hbox{\rm I\kern-.18em F}} 
\def\Naturals{\hbox{\rm I\kern-.17em N}}
\def\naturals{\Naturals}
\def\integers{\hbox{\rm Z\kern-.3em Z}}
\def\hh{\hrule height0.9pt width1.1em}
\def\vv{\vrule width0.8pt depth0.12em height0.92em}
\def\square{\vbox{\kern0.15em\hh\kern0.9em\hh\kern-1.05em
            \hbox{\vv\kern0.93em\vv}}}
\def\Lv{L_{\alpha} }
\def\Lvp{L_{\alpha'} }
\def\sm{{\mbox{\boldmath $\sigma$}_m}}
\def\MMv{R_{\mbox{\boldmath $\sigma$}_v}}
\def\tMMv{\widetilde{R}_{\mbox{\boldmath $\sigma$}_v}}
\def\MMm{R_{\mbox{\boldmath $\sigma$}_m}}
\def\tMMm{\widetilde{R}_{\mbox{\boldmath $\sigma$}_m}}
\def\MMj{R_{\mbox{\boldmath $\sigma$}_j}}
\def\sn1{{\mbox{\boldmath $\sigma$}_{n-1}}}
\def\snv{{\mbox{\boldmath $\sigma$}_v}}
\def\eps{\epsilon}
\def\tb{\hspace*{0.5 in}}
\hyphenation{eigen-value so-lu-tions heur-is-tics mod-eled de-note
prob-a-bil-ity start-ing}

\title{Efficient Hashing with Lookups in two Memory Accesses
}

\author{ 
Rina Panigrahy \thanks{Cisco Systems, San Jose, CA 95134. E-mail: {\tt rinap@cisco.com}.}
}

\maketitle
\begin{abstract}

The study of hashing 
is closely related to the analysis of balls and bins. Azar et. al. \cite{ABKU99} showed that instead 
of using a single hash
function if we randomly hash a ball into two bins and place it in the smaller of the two, then this 
dramatically lowers the maximum load on  bins. 
This leads to the concept of two-way hashing where 
the largest bucket contains $O(\log\log n)$ balls with high probability. The hash look up will now search in both the buckets 
an item hashes to. Since an item may be placed in one of two buckets, we could potentially move an item 
after it has been initially placed to reduce maximum load. 
We show that by performing moves during inserts, a maximum load of $2$ can be maintained on-line, with high probability,
 while supporting hash update operations.
In fact, with $n$ buckets, even 
if the space for two items are pre-allocated per bucket, as may be desirable in hardware 
implementations, more  than $n$ items can be stored giving a high memory utilization.
This gives a simple practical hashing scheme with
the following properties:

\begin{itemize}
\item Each lookup takes two random memory accesses, and reads at most two items per access.
\item Each insert takes $O(\log n)$ time and  up to  $\log\log n+ O(1)$ moves, with high probability, and
	constant time in expectation.  
\item Maintains $83.75\%$ memory utilization, without requiring dynamic allocation during inserts.
\end{itemize}

We also analyze the trade-off between the number of moves performed during inserts and the maximum
load on a bucket. By performing
at most $h$ moves, we can maintain a maximum load of $O(\frac{\log \log n}{ h \log(\log\log n/h)})$.
So, even by performing one move, we achieve a better bound than by performing no moves at all.
\end{abstract}

 
\section{Introduction}

The study of hashing 
is closely related to the analysis of balls and bin. One of the classical results 
in this area is that, asymptotically, if $n$ balls are thrown
into $n$ bins independently and randomly then the largest
bin has $(1+o(1))\ln n/ \ln\ln n$ balls, with high probability. Azar et. al. \cite{ABKU99} showed
 that instead of using a single hash
function, if we randomly hash a ball into two bins and place it in the smaller of the two, then this 
dramatically lowers the maximum load on bins. This leads to the concept of two-way hash functions where 
the largest bucket contains $O(\log\log n)$ balls. The hash look up will now search in both the buckets 
an item hashes to. So dramatic is this improvement that it can be used in practice to efficiently implement 
hash lookups in packet routing hardware \cite{BM01}. The two hash lookups can be parallelized by
placing two different hash tables in separate memory components.

Note that since an item may be placed in one of two buckets, we could potentially move an item 
after it has been initially placed to reduce maximum load. While it was known that if 
all the random choices are given in advance, balls could be assigned to bins with a maximum load of
$2$ with high probability \cite{CS97}, we show that this can be achieved on line while supporting hash up date
operations. In fact, even more  than $n$, up to $1.67n$, items can be stored in $n$ buckets, 
with a maximum load of two items, by performing at most $\log\log n+ O(1)$ moves during inserts,
with high probability. Even if the space 
for two items are pre-allocated per bucket, as desirable in hardware implementations to avoid 
dynamic allocation,
this represents only a $16.25\%$ wastage of space - over $83.75\%$ utilization. 
Memory utilization is a crucial issue in several hash implementations, especially
hardware implementations where a large number of memory components consume critical
resources of board space, ASIC pin count and power.
Our algorithm requires a bfs (breadth first search) exploring at most $O(\log n)$ nodes with high probability and 
constant in expectation. Alternatively, to avoid a bfs, we show that
one could simply perform a random walk of length $O(\log n)$ to maintain a maximum load of
two provided $m < 0.65n$; for larger $m$ this would give a constant load as long as  
$m=O(n)$.  

We also analyze the trade-off between the number of moves performed during inserts and the maximum
load on a bucket. A solution requiring fewer moves may be more attractive in practice as moves may be expensive; 
also it may be desirable to avoid a bfs traversal that may be infeasible in hardware 
implementations. By performing at most $h$ moves during inserts,
 we can maintain a maximum load of $O(\frac{\log \log n}{ h \log(\log\log n/h)})$.
So even by performing one move, we achieve a better bound than by performing no moves at all.

This idea of moving items has been used earlier in cuckoo hashing \cite{PR01}, however, they
allow only one item per bucket.  With two hash tables this requires $100\%$ memory overhead.
A probabilistic analysis of this hashing method was done in \cite{DM03} showing that the 
amortized insert time was constant.  Fotakis et al \cite{FPSS03} generalized the method to
$d$-ary hashing, using $d$ hash tables, but still allowing only one item per bucket.  They showed that
with $\eps$ memory overhead, one can support hash lookups in $O(\ln^2 1/\eps)$ probes and 
constant amortized insert time.
But no  high probability bounds on the insert time, nor a trade off between maximum bucket
size and the number of moves required during inserts was known for the prior art.   
Also, in practice, memory operations requiring more
random accesses is more expensive than reading the same amount of memory in few accesses and 
larger bursts. The latency of the initial random access is much higher than
that of fetching data from subsequent locations.
Also, in hardware implementations, probing a large though constant number of tables will require 
as many memory components to be accessed efficiently in parallel. 
Our method involves two memory accesses and achieves a $83.75\%$ memory utilization. Note that this
utilization is what can be provably achieved and is not tight; although we can show an upper bound
of $93\%$ for our algorithm.

Other related work includes the first static dictionary data structure with constant 
look up time by Fredman, Komlos and Szemeredi \cite{FKS84} that was generalized to a 
dynamic data structure by Dietzfelbinger et al. in \cite{DKMMRT94} and \cite{DM90}. In practice,
however, these algorithms are more complex to implement than cuckoo hashing. Extensive
work has been done in the area of parallel balls and bins \cite{ACMR95} and  the related study of
algorithms to emulate shared memory machines (as for example, PRAMs)
on distributed memory machines (DMMs) \cite{DM93} \cite{CMS95}  \cite{MSS96} \cite{SEK00}. 
This setting involves a
parallel game of placing balls in bins (the so-called collision game) where all
n balls participate in rounds of parallel attempts to assign balls to bins.
In each round, you test both locations of every ball that
has not been placed yet. If a ball has a location tested by at most some
constant number of other balls, you place it.
It has been shown in  that $log log n + O(1)$ rounds indeed suffice to place
all n balls, with high probability \cite{DM93} \cite{CMS95}. This however does not imply 
our result that $log log n + O(1)$ moves are sufficient to maintain maximum load of
$2$ because of the different setting. 


\section{Overview of Techniques}\label{sec1}

Viewing buckets as bins and items as balls, we can look at the hashing process as if
$m$ balls are being assigned to $n$ bins. For each ball two bins are chosen at random.
If the bins are imagined to be the vertices of a graph, the two bins for a ball can
be represented by an edge. This gives us a random graph $G$ on $n$ vertices
containing $m$ edges. By making this graph directed, we could use the direction of an edge to
indicate the choice of the bin among the two for placing the ball. The  
direction of each edge is chosen online by a certain procedure. The load of a vertex (bucket)
is equal to its in-degree.  For each edge (item) insertion, the two-way hash algorithm
directs the edge towards the vertex with the lower in-degree. During the hash process, say $U$
is one of the vertices a ball gets hashed to. Observe that if $VU$ is a directed edge, and if 
the load on $V$ is significantly lower, we could perform a move from $U$ to $V$, thus
freeing up a position in $U$. Essentially, in terms of load, the new ball could be added to 
either $U$ or $V$, whichever has a lower load. This principle could be generalized to the case where there is a directed 
path from $V$ to $U$, and would result in performing moves and flipping the directions along
 all the edges on the path.
 If there is a directed sub-tree rooted at $U$, with all edges leading to the 
root, we could choose the least loaded vertex in this tree to incur the load of the new ball. 
With this understanding, we will say that $W$ is a child of $X$ if $XW$ is a directed edge.
So, our hash insert algorithm looks as follows.

\begin{itemize}

\item Compute the two bins  $U_1$ and $U_2$ that the new item to be inserted hashes to. 
\item Explore vertices that can be reached from $U_1$ or $U_2$ by traversing along directed edges in the
	reverse direction.
\item Among such vertices, find one, $V$, with low load that can be reached say from $U_1$.  
\item Add the new item to $U_1$ and perform moves along the path from $U_1$ to $V$
	 so that only the load on $V$ increases by one.

\end{itemize}

Let $s=2m/n$ denote the average degree of the undirected random graph $G$. Note that  the
same graph $G$ can be viewed as a directed or an undirected graph.
Throughout the paper $G$ refers
to the undirected version unless stated otherwise or clear to be so from the context.
Throughout the paper we will 
assume that $s$ is a constant. It turns out that the success of our algorithm in maintaining
low maximum load depends on the absence of 
dense subgraphs in this random graph.  We show that such dense subgraphs are absent when
$s < 3.35$, giving an algorithm that works with bucket size at most $2$ and requiring at most
$\log\log n+ O(1)$ moves for inserts with high probability (section \ref{seccon}).
Note that the bound of $3.35$ for
$s$ may not tight but is provably no more than $3.72$. We then analyze the trade off between 
number of moves during inserts and maximum bucket size using the technique of witness trees
\cite{CMS95} \cite{MSS96} \cite{ACMR95},
 making significant adaptations to our problem (section \ref{sectrade}).

\section{Constant Maximum Bucket Size}\label{seccon}

In this section we show that  for $s<3.35$ by performing at most $\log\log n+ O(1)$ moves, 
we can ensure that with high probability no bucket gets more than $2$ items.

For an insert, we search backwards from a given node in bfs order, traversing directed edges
in reverse direction,  looking
for a node with load at most one. To simplify the analysis, we assume that during the backward 
search, the algorithm visits only $2$ children for each node even if more may be present.
 We will show that 
by searching to a depth of $\log\log n+ O(1)$, with high probability, we find a node with 
load at most one. 
First, we show that if the backward search is allowed to proceed to unlimited depth, 
the success of the algorithm is related to a certain property of the random graph $G$.

\begin{lemma}\label{lemdense}
If the backward search during inserts is allowed to proceed to any depth, the above algorithm succeeds in inserting all $m$
items while maintaining a maximum load of $2$ if and only if the graph $G$ does not have
a subgraph with density greater than $2$. Here density is the ratio of 
number of edges to vertices in the subgraph. 
\end{lemma}  
\begin{proof}
Clearly, if there is such a subgraph, it is impossible to orient the edges so that the in-degree
on every vertex in the subgraph is at most $2$. So it is not possible to have inserted all elements
and still have a load at most $2$ on every vertex.

Conversely, if an insert does not succeed, it means the backward search does not find a node of load less than $2$.
Since the search was not limited to a bounded depth, it must have got stuck in a set of nodes
all with load at least two and 
leading to each other by traversing edges in reverse direction.  Then this set of nodes is a subgraph of density at least two.
\end{proof}

The existence of dense subgraphs in random graphs displays a {\em critical point} behavior;
that is, there is a sharp threshold such that almost all random graphs with edge-density
 larger than the threshold value have such a subgraph
and almost all with edge-density less than the threshold value have none.
This is because the existence of a dense subgraph is
a monotone property, and all such properties were shown to display a sharp threshold behavior
 by Friedgut  and Kalai \cite{FK96}. A closely related property, the existence of 
a $k$-core in random graphs, has been studied extensively and the threshold values have been pinned down exactly.  
A $k$-core is a maximal non-empty subgraph where every node has degree at least $k$.  Pittel et al \cite{PSW96}
showed that for the existence of a $3$-core the critical value is about 3.35. Note that existence
of a subgraph with density greater than $2$ implies existence of a $3$-core.  This is because by iteratively
deleting nodes with degree at most $2$ we must be left with a non-empty $3$-core as 
the number of deleted edges is at most than twice the number of deleted vertices, less than the
total number of edges. This means
that the threshold value for the existence of a $2$-dense subgraph is at least $3.35$. We will show that
it lies between $3.35$ and $3.71$.  Further, we will show that for $s \le 3.35$, not only does 
an inserts succeed with high probability but also takes less than $\log\log n+ O(1)$ moves.  
It is interesting that this value of $s$ coincides with the threshold value for existence of a $k$-core,
but not surprising as we use methods similar to that for $k$-core in lower bounding
the threshold value. Although this value was also shown to be tight for the existence of $3$-core
 by Pittel et al \cite{PSW96}, it is unlikely to be so for the existence of $2$-dense subgraph.

Since our strategy is to search for a node with load at most one, first we show that it is unlikely
to get stuck in a situation where $o(n)$ nodes have been explored, each with load at least $2$, 
and they all lead to one another with no new nodes to visit.  This follows from the
following lemma as if we do get stuck, we have found an induced subgraph where every node has
in-degree at least two.

\begin{lemma}\label{lemstuck}
With high probability, $1-O(1/n^2)$ there does not exist an induced subgraph of size $o(n)$
in $G$ where every node has in-degree at least $2$. This implies that the backward search
cannot get stuck with high probability if it is allowed to proceed to any depth.
\end{lemma}  
\begin{proof}
If there is such a subgraph of $x$ nodes, it must have at least $2x$ edges.
We will show that the probability of such an event is negligible.
Number of ways of choosing $x$ vertices and $2x$ edges from the $m$ edges is 
${n \choose x}  {sn/2 \choose {2x}} $. Probability of a given edge falling in this subgraph 
is $\frac{{x\choose 2}}{{n\choose 2}} \le \frac{x^2}{n^2}$
So the total probability is 
\beq
	& \le & {n \choose x}  {sn/2 \choose {2x}} (\frac{x^2}{n^2})^{2x} \\
	& \le & (\frac{en}{x})^x (\frac{esn/2}{2x})^{2x} (\frac{x}{n})^{4x} \\
	& \le & ( \frac{e^3 s^2 x}{16n} )^x
\eeq
Since $x$ is at least $2$ and at most $o(n)$, this probability is $O(1/n^2)$ 
\end{proof}

Let us perform a bfs on the undirected graph $G$ starting from a certain node $V$ to a depth of $h$.
Note that this is different from the backward search from the same node to a depth of $h$ that 
also involves a bfs along directed edges in reverse direction.  To distinguish between the two
we will refer to the former as `bfs on the undirected graph' and the latter as `backward search'.
Let $BFS_h(V)$ denote the subgraph visited by the bfs on the undirected graph to a depth of $h$. 
Clearly the nodes visited in the backward search to a depth of $h$ will be a subset of
those visited in $BFS_h(V)$ to a depth of $h$.
 We will compare this bfs on the random undirected graph $G$ to a branching process.  
Since $sn/2$ edges are randomly thrown into the graph $G$ on  $n$ vertices, each of the 
total of $sn$ endpoints of these edges are chosen randomly. If
we ignore the possibility of forming self loops and  choose these endpoints independently,
a node will have $k$ edges incident on 
it with probability $\alpha_k = {sn \choose k} (1/n)^k (1-1/n)^{sn-k}
\approx e^{-s} s^k / k!$ (accurate for large $n$ and $k << n$ and can be safely used in summations). This probability is asymptotically accurate
even if we condition on a certain subgraph with at most $o(n)$ nodes and edges as it makes
a negligible difference in the ratio of remaining nodes and edges.

Consider a branching process where each 
node has $k$ children with this probability $\alpha_k$; this branching process is completely separate from the bfs
and simply constructs a tree where each node has $k$ children with this probability $\alpha_k$.
Let $BRT_h$ be the tree obtained by running such
a branching process to a depth of $h$.  
We will later show that assuming no cycles are found during the bfs to depth $h$, the tree $BFS_h(V)$ 
that is obtained has asymptotically the same distribution as that of $BRT_h$.  
If $BFS_h(V)$ is a tree and only contains  nodes with load at least two, then one can
embed a complete, balanced binary tree of depth $h$ in it.
We will show that the probability of this event is close the probability of the being able to
embed a complete, balanced binary tree of depth $h$ in $BRT_h$.  The next two lemmas show
that it is unlikely to be able to embed a complete, balanced binary tree of depth $h$ in $BRT_h$
if $s \le 3.35$.

\begin{lemma}\label{lempi}
Let $p_i$ be the probability that a complete, balanced binary tree of depth $i$ can be embedded
in the tree $BRT_i$ obtained by running the branching process to depth $i$.  
Then $p_{i+1} = 1 - e^{-p_i s} (1+p_i s)$ 
\end{lemma}  
\begin{proof}
We compute $p_i$ recursively.  Look at a node at height $i+1$. At least two of its children
must satisfy the recursive property which happens with probability $p_i$.  If there are $k$
children, probability that less than $2$ of them satisfy the property is 
$(1-p_i)^k + k p_i (1-p_i)^{k-1}$. Probability of having $k$ children $= \alpha_k$

So, 
\beq  
p_{i+1} &=& \sum_{k\ge 2} \alpha_k (1 - (1-p_i)^k - k p_i (1-p_i)^{k-1}) \\
	&=& \sum_{k\ge 2} \frac{ e^{-s} s^k}{k!} (1 - (1-p_i)^k - k p_i (1-p_i)^{k-1}) \\
	&=& 1 - e^{-s}e^{s(1-p_i)} - e^{-s}p_i s e^{s(1-p_i)} \\
	&=& 1 - e^{-p_i s}(1+ps)
\eeq
\end{proof} 

\begin{lemma}\label{lemloglogn}
For any $s \le 3.35$, the probability that a complete, balanced
binary tree of depth $h$ can be embedded in $BRT_h$,  can be made  smaller than $1/n^c$, for
any constant $c$, by choosing $h= \log\log n + O(1)$ 
\end{lemma}  
\begin{proof}
As long as $s$ is such that $1 - e^{-p s} (1+p s)$ is always less than $p$ for any $p \in (0,1]$, 
the sequence $p_i$ is monotonically decreasing. If $ps$ is very small this expression is close to
$p^2s^2$ as $e^{-ps}$ can be approximated as $1-ps$. In a constant number of steps $p$ can be made
smaller than $1/(10s)$, after which it starts decreasing quadratically each step 
with the recursion $p_{i+1} = p_i^2 s^2$ that is equivalent to $p_{i+1}s^2 = (p_i s^2)^2$.  
So after this point, in $\log\log n + O(1)$ steps, the probability should drop below $1/n^c$.

We want that for any $p \in (0,1]$
\beq
 1 - e^{-p s} (1+p s) < p \Leftrightarrow  e^{p s} < \frac{1+p s}{1-p} 
\eeq
By writing both sides as a Taylor series in $p$ and comparing, we see that this is satisfied if
\beq
 s^2/2 < s+1 \Leftrightarrow   s  < \sqrt 3 +1 < 3.74
\eeq
A better value of $3.35$ is obtained by plotting graphs for the functions $f(x) = 1+xs - e^{x s} (1-x)$
 in the interval $[0,1]$ showing that 
$f(x) >0 $ for $s \le 3.35$ in this interval.
\end{proof}

This value of $s$ is tight; that is, for $s>3.36$ it can be shown that $p$ converges to $0.5$,
implying that it is possible to embed a binary tree.

Next we extend this result on the tree obtained from the branching process to any tree that
may be obtained by the bfs.

\begin{lemma}\label{lembfstree}
With high probability, $1-O(1/n^{c-1})$, there does not exist a node $V$ in $G$ so
that the bfs from $V$ to depth $h = \log\log n + O(1)$ does not encounter any cycles and results in
a tree containing a complete, balanced
binary tree of depth $h$ embedded in it. (Note that the bfs could be
performed from an edge $UV$ where the first level of bfs 
from the root $V$ does not visit $U$. This is a technical detail that will be used later.)
\end{lemma}  
\begin{proof}
We will argue that if the bfs results in a tree, its distribution is asymptotically same as that produced
by the branching process. First note that the total number of nodes visited is small as compared to
$n$, as the maximum degree $d$ is $O(\log n)$ with high probability  and the values of $h$ in consideration
is $O(\log\log n)$, and so the total number of nodes, $d^h$, is $(\log n)^{O(\log\log n)}$.

Even if we condition on the existence of a certain subgraph with at most $o(n)$ 
nodes and edges it makes
a negligible difference in the ratio of remaining nodes and edges. So during the bfs,
after exploring say at most $x$ nodes and edges ($x$ is at most $(\log n)^{O(\log\log n)}$), 
the conditional probability that the next node to be expanded
 has $k$ ($k$ is at most $O(\log n)$) edges emanating from it all of 
which lead to new nodes, is very close to $\alpha_k$.
It can be verified that the conditional probability is at most 
${n \choose k} {sn/2 \choose k} k! (\frac{2}{(n-x-1)^2})^{k} 
{(1 - \frac{2(n-x)}{(n-1)^2})}^{sn/2-x}$ -- number of ways of choosing k child nodes and edges
to those nodes is at most ${n \choose k} {sn/2 \choose k} k!$; probability that
one of the k edges leads to the chosen child is at most $1/{n-x \choose 2}$; probability
that each of the remaining $sn/2 -x$ edges are not incident on this node is at least
$(n-x)/{n \choose 2}$ as at least $n-x$ edge positions are forbidden. This upper bound
differs from $\alpha_k$ by at most a multiplicative factor of 
$1+O(kx/n)$, for the small values of $k$ and $x$ under consideration. So the probability
that the bfs and the branching process produce identical trees of a given structure with at most
$x$ nodes, differ by at most a multiplicative factor of $1+O(kx^2/n) = 1 + o(1)$.
So by applying this argument to all possible trees that can have a complete, balanced binary tree of depth $h$,
embedded in it, we can conclude that since with high probability of $1-O(1/n^{c})$,
$BRT_h$ cannot have a complete binary tree embedded
in it, same must be true about $BFS_h(V)$ even if it were a tree. Clearly this can be extended
to all vertices $V$ with high probability of $1-O(1/n^{c-1})$.
\end{proof}

So far we have only considered the case that $BFS_h(V)$ is a tree. Let us prove that 
it is very unlikely that the bfs finds too many edges that create cycles, where
by cycle-creating edges we mean the edges that lead to already visited nodes during the search..

\begin{lemma}\label{lemcycle}
With high probability, $1-O(1/n^c)$ there does not exist a subgraph of $x \leq c\log n$
nodes in $G$ with at least $x+O(c)$ (precisely, $x + c(4+\log(s/2))$) edges.
\end{lemma}  
\begin{proof}
If there is such a subgraph of $x$ nodes, we will show that the probability 
of such an event is negligible.
Number of ways of choosing $x$ nodes and $x+u$ edges from the $sn/2$ edges is 
${n \choose x}  {sn/2 \choose {2+u}} $. Probability of a given edge falling in this subgraph 
is $\frac{{x\choose 2}}{{n\choose 2}} \le \frac{x^2}{n^2}$
So the total probability is 
\beq
	& \le & {n \choose x}  {sn/2 \choose {x+u}} (\frac{x^2}{n^2})^{2x+2u} \\
	& \le & (\frac{en}{x})^x (\frac{esn/2}{x+u})^{x+u} (\frac{x}{n})^{2x+2u} \\
	& \le & (\frac{es}{2})^u e^{2x} (\frac{s}{2})^{x} (\frac{x}{n})^{u} \\
	& \le & (\frac{es}{2})^u e^{2c\log n} (\frac{s}{2})^{c\log n} (\frac{c\log n}{n})^{u} \\
	& \le & (\frac{es}{2})^u n^{c(2+\log(s/2))} (\frac{c\log n}{n})^{u}
\eeq
By setting $u=c(4+\log(s/2))$ this becomes $O(1/n^c)$
\end{proof}

The following lemma shows that a bfs to a depth of $o(\log n)$ can not encounter
more than $5c$ edges that create cycles.

\begin{lemma}\label{lemmst}
For $s\le 3.35$, with high probability, $1-O(1/n^c)$, in a subtree $T$ of $G$ with height $o(\log n)$ 
there cannot be $5c$  edges in $G$  that are between nodes in $T$ but are not edges of $T$.
\end{lemma}  
\begin{proof}
For if there were, then consider the tree spanning end-points of these $5c$ edges 
from $G$ not in $T$, obtained by taking the union of all the paths from these end-points to the root.
As the number of endpoints of these $5c$ edges is
at most $10c$ and each requires at most $o(\log n)$ edges to connect to the root,
 the size, $x$, of this spanning tree is clearly less than $c \log n$. 

Adding the $5c$ edges to the spanning tree gives us at least $x+5c$ edges. 
By lemma \ref{lemcycle}, for $s<4$, this is unlikely and has probability 
at most $O(1/n^c)$.
\end{proof}

Now we will show that a large, complete binary tree cannot be embedded in $G$.
\begin{lemma}\label{lemnobinary}
With high probability, $1-O(1/n^c)$, it is not possible to embed a complete, balanced
binary tree $B$ of height $h = \log\log n + O(1)$ in the random graph $G$.
\end{lemma}  
\begin{proof}
Assume that we can embed such a binary tree $B$ rooted at $V$ in $G$. 
Perform a bfs from $V$ to a depth of $h$. By
lemma \ref{lemmst},
 at most $5c$  cycle creating edges can be found with high probability in $BFS_{h}(V)$.
Let $BFS'_{h}(V)$ denote the tree obtained by deleting these $5c$ edges from $BFS_{h}(V)$.
There must be some node $V'$ in $B$ at depth 
at most $\log(5c)+1$ so that the binary subtree rooted at that node is still intact in 
$BFS'_{h}(V)$; that is it does not contain any of the $5c$ deleted edges. 
Let $B'$ denote the binary subtree of $B$ rooted
at $V'$. Now look at the at most $10c$ paths from the endpoints of these deleted edges
to $V$. Since any single path can intersect at most $2$ nodes at a certain level in $B'$,
there must be some node $V''$ at depth $\log(20c)+1$ in $B'$ that is not on any of
these $10c$ paths. Also, at least one of the two children of $V''$ in $B'$  (say $W$) must also be 
a child of $V''$ in $BFS'_{h}(V)$, as $V''$ has at most one parent in $BFS'_{h}(V$). 
Look at the binary subtree $B''$ of $B'$ rooted at $W$. The height of $B''$ differs
from that of $B$ by at most $\log(5c) + \log(20c) + 3$. Also the bfs from the edge $V''W$ 
(that is, the first level of the bfs from $W$ does not visit $V''$)
is free of cycles as otherwise $V''$ is on one of the $10c$ paths. Further it has
a complete, balanced binary tree $B''$ embedded in it. By choosing $h$ large enough we can
ensure that height of $B''$ is at least that required by lemma \ref{lembfstree} giving a contradiction
\end{proof}

We are now ready to prove that during an insert a backward
search to a depth of $\log\log n + O(1)$  must find
with high probability a node with load less than $2$. The total search time
is at most $O(\log n)$.  

\begin{theorem}\label{thmc1}
For $s \le 3.35$, with high probability, $1-O(1/n^2)$, during an insert,
if we traverse backward to a depth of
$\log\log n + O(1)$, we will have found a node with load less than $2$, with high probability,
while searching 
at most $O(\log n)$ nodes. The expected time for this search is $O(1)$.
\end{theorem}
\begin{proof}
Assume that during an insert, we don't find a node of load less than $2$. 
Then since with high probability
by lemma \ref{lemstuck} we cannot get stuck after a few levels and by lemma \ref{lemmst} we cannot encounter more 
than $5c$ cycle producing edges, there must be a node at depth $\log(5c)+1$ so that
the backward search under that does not find any cycles. This gives a complete binary tree 
of height $\log\log n + O(1)$, contradicting lemma \ref{lemnobinary}.

The expected depth of search is constant as can be seen by the
quadratic drop of $p_i$ with $i$.
\end{proof}

This proves  that inserts can be made while maintaining a maximum load of $2$, with high probability.
The algorithm works even if the number of items, $m$ is greater than $n$  
 as long as $2m/n \le 3.35$. Even if the two entries in each buckets are statically allocated,
we can achieve a 
memory utilization $m/(2n)$ of  $3.35/4 > 83.75\%$.  Thus the memory wastage is only $16.25\%$. 

Note that our value of $s = 3.35$ may not be tight for maintaining a maximum load of two as 
the calculation was done based on existence of a complete binary tree, which may not  be
necessary for the existence of a $2$-dense subgraph nor for being able to perform inserts in 
$\log\log n + O(1)$ moves.
It is easy to show, however, that for $s > 3.72$, it is impossible to maintain
a maximum load of two.  This is because for such a random graph, by deleting isolated nodes
and nodes of degree one, we end up with a non-empty component with density greater than $2$.

{\it Generalizing to constant bucket size larger than 2}:
Our analysis for maximum bucket size of 2 can be generalized to any constant maximum load $i$. 
It turns out that the best provable memory utilization remains around $80\%$ for initial value
of $i>2$ and then drops for larger $i$.

\subsection {Random Walk}
The previous algorithm performs a bfs.  An alternate algorithm is to simply perform a
random walk to look for a lightly loaded node.  We  show that for $m < 0.65n$, a random walk
of length $O(\log n)$ will reveal a node with load at most $1$.
We omit the proof for lack of space.

\begin{theorem}\label{thmr1}
With high probability, $1-O(1/n^2)$, for any $s < 1.3$, a random walk of length
$O(\log n)$ will find a node with load at most two.
\end{theorem}

\begin{theorem}\label{thmr2}
With high probability, $1-O(1/n^2)$, for $m=n$, a random walk of length
$O(\log n)$ will find a node with load at most $4$.
\end{theorem} 
 

\section{Generalizing to fewer moves}\label{sectrade}

So far we have looked at the number of moves required to maintain a constant load.
Here we examine the maximum load when fewer bins
are explored. In particular, we could examine only the two bins and their
children.  So, if
an item gets hashed to say $U_1$ and $U_2$, we could examine only $U_1, U_2$ and the 
children of $U_1$ and $U_2$, and pick the least loaded of these to bear the new load. This 
would require at most one move. Instead of examining the children to a depth of one, 
we could explore all the descendants to a depth of $h$ by performing a bfs along directed edges
in the reverse direction. By restricting
the search to a depth of $h$, we ensure that at most $h$ moves are required.
In this section we upper bound the maximum load when all descendants up to depth $h$
are examined during inserts.

The basic intuition is that if the load of the new item is borne by a node with 
load $i$, then each of examined nodes must have at least $i$ children. So we must
have explored roughly a total of $i^h$ nodes, each with a load of at least $i$. If 
$p_i$ is the probability of a node having load at least $i$, then assuming these 
events are independent, they happen with probability $p_i^{i^h}$. This gives us
approximately, $p_{i+1} = p_i^{i^h}$, and so $p_i = 2^{-\Omega{(i-1)!^h}}$. $p_i$ becomes
$o(1/n^c)$ for $i > O(\frac{\log \log n}{ h \log(\log\log n/h)})$
We give a more formal proof of this result without making the independence assumption.

Our proof is based on the {\em witness tree}
approach -- one of earliest uses of this approach can be found 
 in \cite{CMS95} \cite{MSS96} \cite{ACMR95}.  Consider an event that
 leads to a load of
$6l$ at a certain node.  For this event to happen,  we will show that there must exist a tree of large size
obtained by tracing all the events that must have happened earlier.  The approach however requires significant adaptation to our problem
as the directions of the edges change over time. To simplify the exposition, we will state
the proof assuming $m=n$ ($s=1$); essentially, the same proof works for any constant $s$.

{\em Construction of the witness graph:}
 Whenever the load of a node $X$ becomes $i$, there must be a unique edge
whose insertion causes this to happen.  Say $U_1 U_2$ was this edge; that is, $U_1$ and 
$U_2$ are the bins to which the item got hashed. Look at the directed graph when this edge
was being added.  During the insertion, a backward search to a depth of $h$ was performed
from both $U_1$ and $U_2$. Say the node $X$ was obtained
by traversing back from $U_1$ to depth of at most $h$.  We will say that the edge $U_1 U_2$
is the {\em $ith$ contributing-edge} of $X$, $U_2$ is the {\em $ith$ contributing-peer} of $X$, and the
directed path from $X$ to $U_1$ along which moves were made, is the {\em $ith$ contributing-path} for $X$. 
Since $X$ is a node with minimum load among the ones visited,
it must be the case that all the nodes
at depth at most $h$ from $U_2$ must have load at least $i-1$. 
 Note that the contributing edge
$U_1 U_2$ must be newer than and therefore distinct from all the edges traversed in 
backward search from $U_2$.
Also for each node and each value of $i$ the $ith$-contributing edge has to be unique.

The witness graph is obtained by recursively
chasing contributing edges for nodes visited in the backward search from the 
contributing peer $U_2$. First we make a simplifying assumption that
during the construction of the witness graph, we never run into  cycles, always 
leaving the graph as a tree. Later, as in section \ref{seccon}, we will
argue that the number of edges that produce cycles is few enough that they can be ignored.
Our goal is to obtain a large witness tree with high degree nodes and argue that such a
subgraph is unlikely to exist in $G$. The problem is that even with our assumption of
not encountering cycles, it is still possible to visit earlier nodes through contributing
paths, as contributing paths could completely consist of edges in the visited subtree, 
not leading to any new nodes. We overcome this issue by computing  
the witness tree by the following  recursive procedure.\\

For a  given $i$-contributing edge $U_1 U_2$:
\begin{itemize}

\item	Say, $U_2$ is the $ith$ contributing-peer  corresponding to this edge; that is, the
	load for this insert was taken by some node under $U_1$.  
	
\item	Look at the subtree, $T$, (must be a tree by assumption of not encountering cycles) 
	obtained by performing a backward search  to depth $h$ from node $U_2$ when the insertion took place. 
	Look at the set of leaves, $L$, of this subtree. At that time each node in $T$ has 
	load at least $i-1$. Since each internal node in $L$ has at least $i-1$ children the
	number of edges in $L$ is at most $2|L|$.

\item	Look at the set $S$ of all edges that are  $j$-contributing edges for some node in $L$, for
	 either $j = i-1$ or $i-2$ or $i-3$. Essentially an edge $e \in S$ if and only if there is 
	a node $V \in L$ and a $j \in \{i-1, i-2, i-3\}$ such that $e$ is the $j$-contributing edge
	for $V$. Since the subtree $T$ has at most $2|L|$ edges and the set $S$ has $3|L|$ edges,
	there must be a set $Q$ of at least $|L|$ edges in $S$ that are outside the subtree $T$.  
	As all the contributing paths leading to these edges are older than the edge $U_1 U_2$
	that connects the subtree to the rest of the witness tree, and since by assumption no
	cycles are encountered, these edges in $Q$ must be outside the entire witness tree constructed
	so far. To avoid cycles, the corresponding contributing paths must branch off $T$
	before reaching the contributing edge in $Q$.
 
\item	Repeat recursively for each $j$-contributing-edge in $Q$, where $j \ge i-3$
	 
\end{itemize}

We chop the recursion depth down to $l$. Also, during the backward bfs, for each node, we pick
 only $l$ children even if more may be present.  Essentially, the witness tree looks like 
a tree of sub-trees linked by contributing paths. Each subtree has $l^h$ ``children'' subtrees 
and no node or edge is repeated.  The height of this tree in terms of number of subtrees is 
$l$. View all edges in this tree as undirected.

For large enough $l$, we will show that 
such a witness tree cannot exist with high probability.  

\begin{lemma}\label{lem2.1} 
Assuming no cycles are encountered while constructing the witness tree, probability that 
such a witness tree exists for $l > \frac{\log \log n}{ h \log(\log\log n/h)} + O(1)$ is 
at most $O(1/n^c)$, where $c$ is any given constant.
\end{lemma}  
\begin{proof}
We will calculate the probability by multiplying the the total number of possible such
trees with their individual probabilities.
Note that all vertices, except those on the contributing paths, have at least $l$ children. 

{\it Ways of choosing $l$ children:}
For a given node, number of ways of choosing these children
is $n \choose l$; number of ways of assigning edges is at most $n^l$; and the
probability of realizing an assignment of edges is at most $(\frac{2}{n^2})^l$.
So the total probability of a given node having $l$ children
is ${n \choose l} n^l  (\frac{2}{n^2})^l < (\frac{2e}{l})^l$.

{\it Ways of choosing a contributing path:}
As for the other nodes, these can only be on contributing paths of length at most $h$
from a node to its contributing-peer. As pointed earlier, all such contributing paths 
of length at most $h$ must branch off the subtree they originate from. For a given contributing
path this branching off point can be chosen in at most $l^h$ ways. 

Number of ways of choosing the rest of the path of length at most $h$ weighted 
by probability  
$\le$ (number of ways of choosing $h$ vertices) $\times$ (number of ways of choosing $h$ edges)
	$\times$ (probability of these edges falling in the right place)
$\leq n^h n^h (\frac{2}{n^2})^h
\leq 2^h$  
So, total number of ways of choosing a contributing path weighted by probability is at most $2^h l^h = (2l)^h$.

{\it Total probability:}
Each subtree has at least $l^{h-1}$ nodes that have $l$ children each, and each subtree 
is rooted at one contributing path.
So number of ways of choosing each subtree weighted by probability is $(\frac{2e}{l})^{l l^{h-1}} (2l)^h \leq  (\frac{4e}{l})^{l^h}$.

Total number of such subtrees is at least $l^{(l-1)h}$.
So total number of ways of choosing witness trees weighted by probability is $(\frac{4e}{l})^{l^h l^{(l-1)h}}
= (\frac{4e}{l})^{l^{lh}}$.
We need to choose $l$ such that this probability is $o(1/n^c)$. This is achieved by 
setting $l$ to $\frac{\log \log n}{ h \log(\log\log n/h)} + O(1)$ 
\end{proof}

So far we have assumed that the construction of the witness graph does not encounter 
any cycle producing edges. We will prove that 
it is very unlikely that it has too many edges that lead to cycles. Again as in 
section \ref{seccon}, using lemma \ref{lemmst} we argue that instead of
starting with a node of load $6l$, if we
start with a node of load $6l+5c$  and attempt to construct the witness tree 
to a recursion depth of $l+1$, it is
very unlikely to encounter more than $5c$ cycle producing edges. 
Since the node under the root contributing-peer has more than $5c$ children, at least
one of them must be such that the witness graph construction under that node is free of cycle-producing
edges, giving the desired result.
  This proves the following theorem.   

\begin{theorem}\label{thm1}
By searching to a depth $h$, with high probability, $1-O(1/n^c)$, 
an insert will not lead to a load of more than
$6\frac{\log \log n}{ h \log(\log\log n/h)}+O(1)$, for any constant $c$.
\end{theorem}

\section*{Acknowledgments} 
I would like to thank Tomas Feder, Michael Mitzenmacher, Christian Scheideler
 and Rajeev Motwani for 
useful discussions. I also wish to thank Artur Czumaj for providing me with a 
draft of his paper \cite{CRS03}.

\end{document}